\documentclass[twocolumn,10pt,letterpaper]{article}
\usepackage{amsopn}
\usepackage{usenix}
\usepackage{graphicx}
\usepackage{caption}
\usepackage{subcaption}
\usepackage[binary,squaren]{SIunits}
\usepackage{tikz}
\usepackage[hyphens]{url}
\usepackage{tabularx}
\usepackage[sort,numbers]{natbib}
\usepackage{enumitem}
\usepackage{amssymb}
\usepackage{multirow}
\usepackage{listings}
\usetikzlibrary{patterns}
\usetikzlibrary{shadows}
\usetikzlibrary{arrows}
\usetikzlibrary{matrix}
\usetikzlibrary{positioning}

\DeclareMathAlphabet{\mathcal}{OMS}{cmsy}{m}{n}

\usepackage{etoolbox}
\apptocmd{\sloppy}{\hbadness 10000\relax}{}{}

\usepackage[small,compact]{titlesec}

\setlist{nolistsep,leftmargin=*}

\newlength\imageheight

\usetikzlibrary{calc}
\usetikzlibrary{matrix}

\tikzset{ 
    table/.style={
        matrix of nodes,
        row sep=-\pgflinewidth,
        column sep=-\pgflinewidth,
        nodes={
            rectangle,
            draw=black,
            align=center
        },
        minimum height=1.5em,
        text depth=0.5ex,
        text height=2ex,
        nodes in empty cells,
        every even row/.style={
            nodes={fill=gray!20}
        },
        column 1/.style={
            nodes={text width=2em,font=\bfseries}
        },
        row 1/.style={
            nodes={
                fill=white,
                text=black,
                font=\bfseries
            }
        }
    }
}

\DeclareMathOperator{\mediator}{mediator}

\begin{document}
\title{Warp: Lightweight Multi-Key Transactions for Key-Value Stores}
\author{\textnormal{Robert Escriva$^\dagger$, Bernard Wong$^\ddagger$, Emin G\"un Sirer$^\dagger$} \\
$^\dagger$\ Computer Science Department, Cornell University\\
$^\ddagger$\ Cheriton School of Computer Science, University of Waterloo
}

\maketitle

\begin{abstract}
Traditional NoSQL systems scale by sharding data across multiple servers and by
performing each operation on a small number of servers.  Because transactions on
multiple keys necessarily require coordination across multiple servers, NoSQL
systems often explicitly avoid making transactional guarantees in order to avoid
such coordination.  Past work on transactional systems control this coordination
by either increasing the granularity at which transactions are ordered,
sacrificing serializability, or by making clock synchronicity assumptions.

This paper presents a novel protocol for providing serializable transactions on
top of a sharded data store.  Called acyclic transactions, this protocol allows
multiple transactions to prepare and commit simultaneously, improving
concurrency in the system, while ensuring that no cycles form between
concurrently-committing transactions.  We have fully implemented acyclic
transactions in a document store called Warp.  Experiments show that Warp
achieves 4$\times$ higher throughput than Sinfonia's mini-transactions on the
standard TPC-C benchmark with no aborts.  Further, the system achieves 75\% of
the throughput of the non-transactional key-value store it builds upon.
\end{abstract}

\section{Introduction}

NoSQL systems have become the de facto back-end for modern applications because
they allow unprecedented performance at large scale.  The defining
characteristic of these systems is their distributed architecture, where the
system shards data across multiple servers to improve scalability.  To further
improve scalability, these systems typically avoid cross-server communication,
which makes it difficult to implement ACID transactions.

Yet, distributed transactions require coordination among multiple servers.  In
traditional RDBMSs, transaction managers coordinate clients with servers, and
ensure that all participants in multi-phase commit protocols run in lock-step.
Such transaction managers constitute bottlenecks, and modern NoSQL systems have
eschewed them for more concurrent implementations.
Scatter~\cite{!arvind:scatter} and Google's Megastore~\cite{!corbett:megastore}
shard the data across different Paxos groups based on their key, thereby gaining
scalability, but incur higher coordination costs for actions that span multiple
groups, and require that operations on the same group be sequenced by the same
Paxos instance.  Google's
Spanner~\cite{!author=hsieh:title=spanner:booktitle=osdi} combines traditional
locking techniques with a novel TrueTime API to provide fast read-write
transactions, and lock-free reads.  Many recent systems propose moving pieces of
the transactional programs themselves to the server.
Calvin~\cite{!abadi:calvin} serializes all transaction inputs via a consensus
protocol, and then deterministically executes application code on servers.
Lynx~\cite{lynx} and Rococo~\cite{rococo} break down the transaction into
multiple atomic fragments, and employ static analysis to detect opportunities
for reordering the shipped code components.  Both techniques rely upon a priori
knowledge and analysis of the transactions.

This paper introduces Warp, a NoSQL system that enables efficient multi-key
transactions with ACID semantics.  Warp offers, to our knowledge, the highest
degree of concurrency in a general purpose serializable NoSQL data store.
Further, it achieves throughput far in excess of previous systems, approaching
75\% of the throughput of the system on which it builds.  The key insight that
enables these performance improvements is a commit protocol called {\em acyclic
transactions}, which allows the system to order transactions on-the-fly without
any prior static analysis, and without moving code to the server.

Three techniques, working in concert, enable acyclic transactions to achieve its
high scalability and performance.  First, acyclic transactions reduce
coordination costs by reducing the number of transactions that are ordered to
the minimum necessary to enforce serializability.  Transactions that operate on
disjoint data or whose executions do not overlap in time will incur zero
coordination costs.  Warp orders only those transactions that concurrently
operate on overlapping data, and does so with minimal overhead.

Second, Acyclic transactions achieve scalability by arranging servers into
data-dependent, dynamically-determined chains, where each chain contains,
solely, those servers which store data affected by the transaction.  This
structure, avoids bottlenecks at transaction managers and improves performance
by cutting communication costs.

Finally, acyclic transactions improve performance by allowing multiple
overlapping transactions to proceed in parallel under the right conditions.
Locking protocols inherently limits concurrency, while traditional optimistic
two-phase protocols can only prepare one transaction per key at a time, because
all reads must be validated in the first phase before the second phase may
begin.  In contrast, Warp enables multiple transactions to prepare
simultaneously by taking advantage of the inherent serialization in its
data-dependent chains.

Overall,  this paper makes three contributions.  First, we outline a novel
protocol for providing efficient, one-copy serializable transactions on a
distributed, sharded data store.  Our protocol provides an unprecedented level
of concurrency and scalability without any synchronicity assumptions or static
analysis.  Second, we describe our implementation of the commercially available
Warp key-value store, including the design of the client.  The system has been
fully implemented and provides language bindings for C, C++, Python, Java, Ruby,
Go, and Node.JS.  Third, we show through macro- and micro-benchmarks that Warp
can provide higher throughput than alternative designs, with fewer aborted
transactions.  Specifically, Warp achieves a throughput that is $4 \times$
higher, with $5 \times$ lower latency, than
mini-transactions~\cite{!aguilera:sinfonia:year=2007}, the closest existing
approach, on the TPC-C benchmark.  The system achieves 75\% the throughput of
the underlying non-transactional key-value store upon which Warp builds.

The rest of this paper is organized as follows.  Sections~\ref{sec:arch}
and~\ref{sec:implementation} describe the acyclic transactions protocol and our
implementation of Warp.  Section~\ref{sec:evaluation} evaluates the performance
of Warp.  Section~\ref{sec:background} surveys existing systems and related work
and Section~\ref{sec:conclusion} concludes.

\section{Design}
\label{sec:arch}

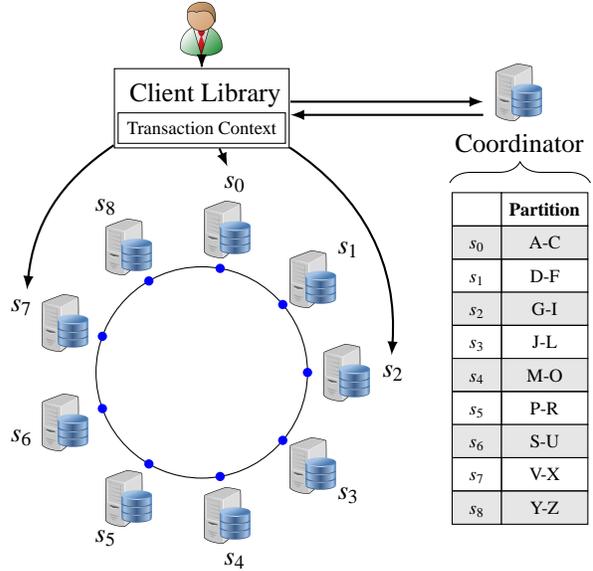
\begin{figure}[t]
\centering
\begin{tikzpicture}
\usetikzlibrary{calc}
\usetikzlibrary{decorations.pathreplacing}
\usetikzlibrary{shapes}

\pgfdeclareimage[height=0.75cm]{dbserv}{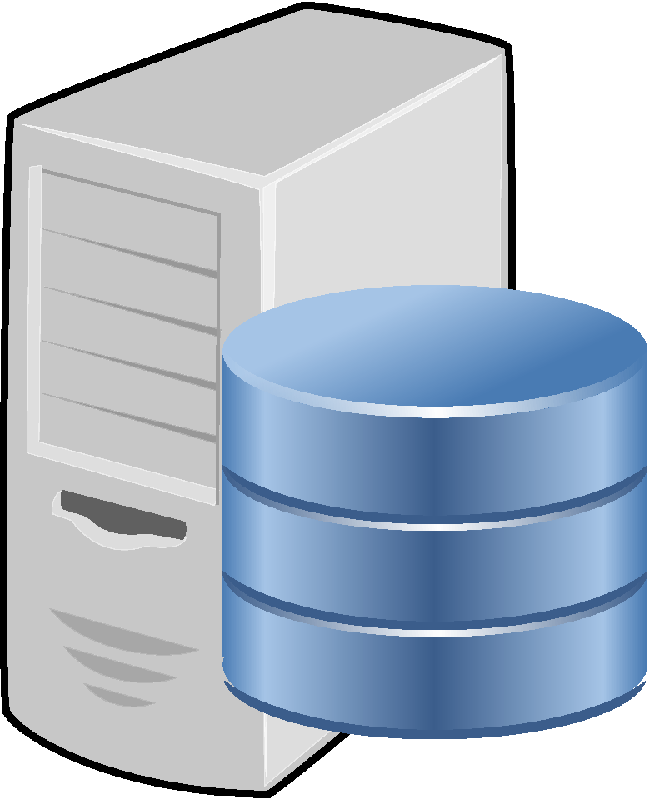};
\settoheight{\imageheight}{\pgfuseimage{dbserv}};
\pgfdeclareimage[height=0.75cm]{client}{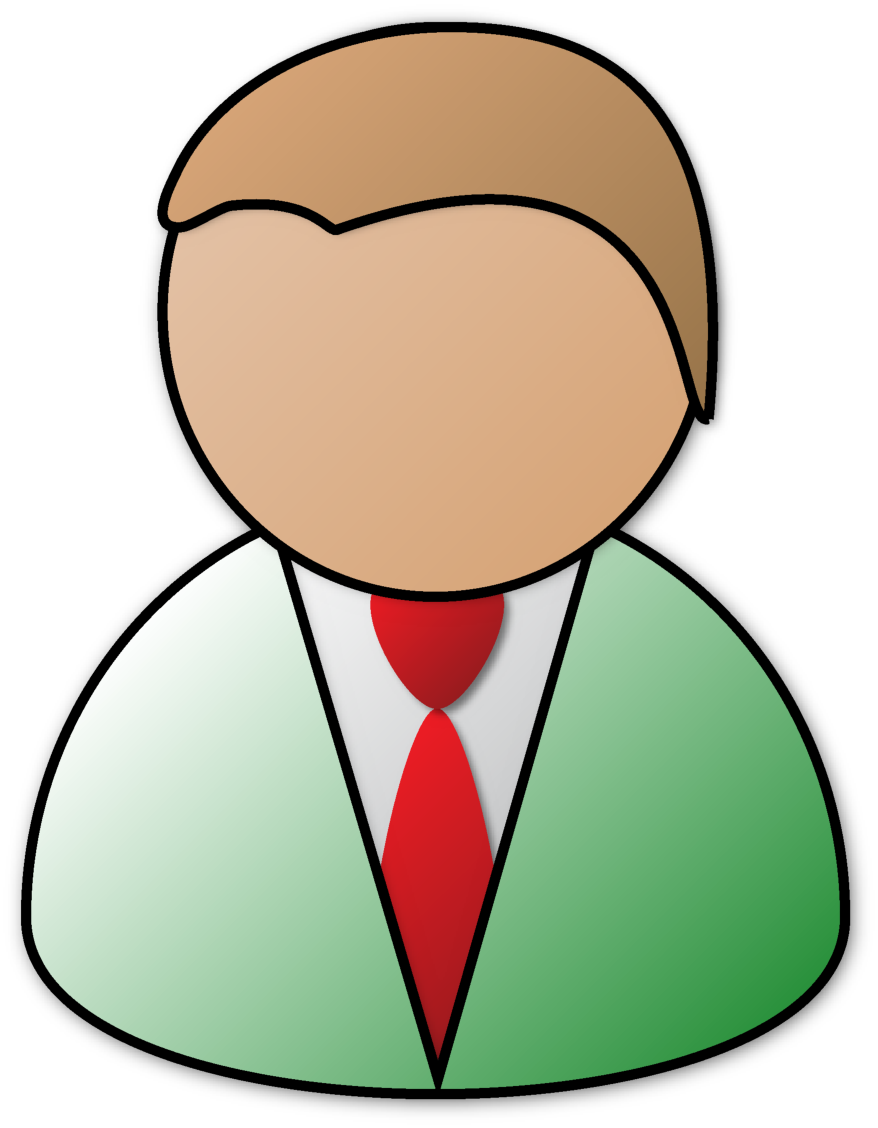};
\settoheight{\imageheight}{\pgfuseimage{client}};


\draw (0em, 0em) circle (4em);

\foreach \name/\angle in {DBServ1/0, DBServ2/40, DBServ3/80, DBServ4/120, DBServ5/160,
                          DBServ6/200, DBServ7/240, DBServ8/280, DBServ9/320}
    \node[circle,color=white,fill=blue,inner sep=0pt,minimum width=0.35em] (dot\name) at (\angle:4em) {};

\foreach \name/\angle in {DBServ1/0, DBServ2/40, DBServ3/80, DBServ4/120, DBServ5/160,
                          DBServ6/200, DBServ7/240, DBServ8/280, DBServ9/320}
    \node[align=center] (\name) at (\angle:5.5em) {\pgfuseimage{dbserv}};

\foreach \name/\angle/\lbl in {DBServ1lbl/0/$s_2$, DBServ2lbl/40/$s_1$,
                               DBServ3lbl/80/$s_0$, DBServ4lbl/120/$s_8$,
                               DBServ5lbl/160/$s_7$, DBServ6lbl/200/$s_6$,
                               DBServ7lbl/240/$s_5$, DBServ8lbl/280/$s_4$,
                               DBServ9lbl/320/$s_3$}
    \node[align=center] (\name) at (\angle:7.25em) {\lbl};


\node[align=center] (client) at (0, 13em) {\pgfuseimage{client}};
\node[rectangle,align=center,below of=client,draw,node distance=3em,minimum height=30pt] (lib) {~Client Library~\\};
\node[rectangle,draw] (ctx) at ($(lib) + (0, -0.75em)$) {\scriptsize Transaction Context};
\draw[->,>=latex,thick] (client) -- (lib);


\node[align=center,right of=lib,node distance=12em] (coordinator)
    {\pgfuseimage{dbserv} \\ \normalsize Coordinator};


\node[right of=lib,node distance=3em,yshift=0.25em] (lc0) {};
\node[below of=lc0,node distance=0.5em] (lc1) {};
\node[right of=lc0,node distance=8em] (lc2) {};
\node[right of=lc1,node distance=8em] (lc3) {};

\draw[->,>=latex,thick] (lc0) -- (lc2);
\draw[->,>=latex,thick] (lc3) -- (lc1);


\draw[->,>=latex,thick] (lib) -- (DBServ3lbl);
\draw[->,>=latex,thick] (lib) edge [bend right=25] (DBServ5lbl);
\draw[->,>=latex,thick] (lib) edge [bend left=30] (DBServ1lbl);


\scriptsize
\matrix (mapping) at (17.2em, .75em) [table, text width=4em]
{
& Partition \\
$s_0$  & A-C \\
$s_1$  & D-F \\
$s_2$  & G-I \\
$s_3$  & J-L \\
$s_4$  & M-O \\
$s_5$  & P-R \\
$s_6$  & S-U \\
$s_7$  & V-X \\
$s_8$  & Y-Z \\
};

\draw[decorate,decoration={brace,amplitude=8pt}] (13.4em, 10.25em) -- (21.0em, 10.25em);

\end{tikzpicture}
\caption{\small Warp's architecture consists of storage servers, the
    coordinator, and the client library.  The coordinator maintains the
    partitioning of the key space across servers, and supplies this mapping to
    the client library.  The client library uses this mapping to directly
    contact storage servers.}
\label{fig:arch}
\end{figure}

Warp builds upon the HyperDex~\cite{!escriva:hyperdex} key-value store by
modifying the existing components to provide transactional guarantees.  Warp's
architecture is based on HyperDex.  To that end, Warp consists of three
components:  the coordinator, clients, and storage servers.  The coordinator
maintains the meta-state for the system, specifically, the partitioning of the
key space across storage servers.  Clients issue requests directly to the
storage servers, where a request may affect a single object, or span multiple
objects.  Each storage server maintains a subset of keys in the system;
collectively, the storage servers hold all data stored by the system.
Figure~\ref{fig:arch} illustrates Warp's overall architecture.

The Warp client library provides isolation by optimistically performing read and
write operations against local state, and verifying that this state remains the
same at commit time.  To perform a read, the library retrieves the requested
data from the storage servers and caches the object within the local
transaction's state, called the {\em transaction context}.  Subsequent reads
within the transaction will be satisfied by the transaction context, if
possible.  Write operations executed within the transaction are not visible on
the servers immediately.  Instead, they are saved to the transaction context to
be written at commit time.  Multiple writes to the same key will overwrite the
locally saved object.  Storage servers are unaware of any modifications written
within a transaction until the client commits the transaction.  To commit the
transaction, the library submits the set of all objects read and all objects
written to the storage servers using the acyclic transactions commit protocol.

\subsection{Commit Protocol}

The acyclic transactions commit protocol processes clients' transactions, and
ensures that they either commit in an atomic, serializable fashion, or abort
with no effect.  The protocol does this for each transaction by simultaneously
validating the values optimistically read by the client library, and ordering
the transaction with respect to other concurrently executing transactions.
While it is relatively easy to validate a transaction by comparing the values
observed by the client to the latest values in the data store, it is
considerably more difficult to order transactions across multiple servers.  The
former is a local check that each server may independently perform during
transaction processing, while the latter requires that multiple servers
coordinate to agree upon the serializable order of transactions.

The key insight of the acyclic transactions protocol is to arrange the servers
for a transaction into a chain, and to validate and order transactions using a
dynamically-determined number of passes through this chain.  Compared to
traditional commit protocols which use fan-out/fan-in communication patterns,
acyclic transactions pass messages forward or backward between adjacent nodes in
the chain.  This ensures that there is at most one server actively processing
each transaction at any one time.  By limiting the parallelism present in a
single transaction, acyclic transactions enable each server to locally make a
binding decision about the fate of the transaction they are processing, and
propagate that decision to the next server in the chain.  Globally, this enables
multiple transactions which modify the same data to execute in
parallel---transactions whose execution other techniques would
serialize---because each pair of concurrent transactions is ordered by exactly
one server that can decide their order without communicating with other servers.
Any decision made by a server will be carried to, and enforced by, the remaining
servers in the chain.

The protocol consists of multiple distinct processes that work in concert to
commit transactions.  To commit the transaction, the client library determines
the chain of servers which process the transaction's commit.  These are {\em
only} those servers that house the data involved in the transaction.  The
servers in this chain follow simple rules to commit the transactions:  a server
passes a transaction forward in the chain only when the server may commit that
transaction; otherwise, the server sends either an abort or a retry request
backward in the chain.  Transactions will be aborted when they fail to validate
the clients' reads, and will be retried to ensure the order between concurrent
transactions is serializable.  When the transaction passes completely through
the chain in the forward direction, the last server in the chain finalizes the
transaction by sending a commit message backwards through the chain.  This
commit message instructs servers to persist the transactions to disk, and to
clean up any transient state related to the transaction.

\subsubsection{Chain Construction}

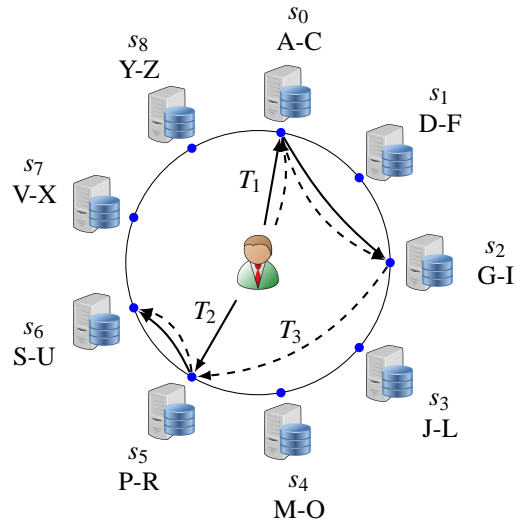
\begin{figure}[t]
\centering
\begin{tikzpicture}
\usetikzlibrary{arrows}
\usetikzlibrary{calc}
\usetikzlibrary{decorations.pathreplacing}
\usetikzlibrary{shapes}

\pgfdeclareimage[height=0.75cm]{dbserv}{figs/database_server.ps};
\settoheight{\imageheight}{\pgfuseimage{dbserv}};
\pgfdeclareimage[height=0.75cm]{client}{figs/business_person.ps};
\settoheight{\imageheight}{\pgfuseimage{client}};


\draw (0em, 0em) circle (5em);

\foreach \name/\angle in {DBServ1/0, DBServ2/40, DBServ3/80, DBServ4/120, DBServ5/160,
                          DBServ6/200, DBServ7/240, DBServ8/280, DBServ9/320}
    \node[circle,color=white,fill=blue,inner sep=0pt,minimum width=0.35em] (dot\name) at (\angle:5em) {};

\foreach \name/\angle in {DBServ1/0, DBServ2/40, DBServ3/80, DBServ4/120, DBServ5/160,
                          DBServ6/200, DBServ7/240, DBServ8/280, DBServ9/320}
    \node[align=center] (\name) at (\angle:6.5em) {\pgfuseimage{dbserv}};

\foreach \name/\angle/\lbl in {DBServ1lbl/0/$s_2$\\G-I, DBServ2lbl/40/$s_1$\\D-F,
                               DBServ3lbl/80/$s_0$\\A-C, DBServ4lbl/120/$s_8$\\Y-Z,
                               DBServ5lbl/160/$s_7$\\V-X, DBServ6lbl/200/$s_6$\\S-U,
                               DBServ7lbl/240/$s_5$\\P-R, DBServ8lbl/280/$s_4$\\M-O,
                               DBServ9lbl/320/$s_3$\\J-L}
    \node[align=center] (\name) at (\angle:9em) {\lbl};


\node[align=center] (client) at (0em, 0em) {\pgfuseimage{client}};

\draw[->,>=latex,thick,solid] (client) to node[left,align=right]{$T_1$} (dotDBServ3);
\draw[->,>=latex,thick,solid] (dotDBServ3) edge [bend right=10] (dotDBServ1);

\draw[->,>=latex,thick,solid] (client) to node[left,align=right,yshift=9pt,xshift=4.5pt]{$T_2$} (dotDBServ7);
\draw[->,>=latex,thick,solid] (dotDBServ7) edge [bend right=15] (dotDBServ6);

\draw[->,>=latex,thick,dashed] (client) edge [bend right=15] (dotDBServ3);
\draw[->,>=latex,thick,dashed] (dotDBServ3) edge [bend right=25] (dotDBServ1);
\draw[-,white] (dotDBServ1) to node[black,below,align=center,yshift=3pt]{$T_3$} (dotDBServ7);
\draw[->,>=latex,thick,dashed] (dotDBServ1) edge [bend left=25] (dotDBServ7);
\draw[->,>=latex,thick,dashed] (dotDBServ7) edge [bend right=35] (dotDBServ6);


\end{tikzpicture}
\caption{\small Clients deterministically construct dynamic chains based upon
    the keys read and written by transactions.  In this example, a client
    submits $T_1$, $T_2$, and $T_3$.  Transactions $T_1$ and $T_2$ operate on
    disjoint keys, $\{k_A, k_H\}$ and $\{k_P, k_T\}$ respectively.  $T_3$
    touches all four keys and forms a chain that includes the chains of
    $T_1$ and $T_2$}.
\label{fig:ex}
\end{figure}

Clients use the transaction's context to construct a chain to commit the
transaction.  To ensure that servers process transactions' operations in a
predictable order, the client library sorts the keys read or written by a
transaction in lexicographical order, and maps this sorted list onto a set of
storage servers.  Because sorting is a deterministic process, transactions with
multiple keys in common pass through their shared set of servers in the same
order.

Figure~\ref{fig:ex} shows how transactions that read and write the same keys
have overlapping chains.  Transaction $T_1$ reads key $k_H$ and writes key
$k_A$, while transaction $T_2$ reads keys $k_P$ and $k_T$.  Transaction $T_3$
writes keys $k_A$, $k_H$, $k_P$, and $k_T$.  The object-to-server mapping
dictates that $T_1$'s chain pass through servers $s_0$ and $s_2$ because these
servers hold objects $k_A$ and $k_H$ respectively.  Similarly, $T_2$ forms a
chain through $s_5$ and $s_6$.  $T_3$ writes the same keys touched by $T_1$ and
$T_2$ and has a chain that passes through the same servers as transactions
$T_1$, and $T_2$.

Constructing chains in this way makes it straightforward to order transactions
that concurrently operate on the same data.  The first server in common between
two transactions' chains can order any two overlapping transactions, and notify
all subsequent servers in both chains.  The mapping maintained by the
coordinator ensures that transactions with data in common will pass through a
common set of same storage servers, even when the mapping is updated to reflect
system membership changes.  Inversely, when two chains do not overlap, there is
no need to directly order their transactions, because they necessarily operate
on disjoint data.

\subsubsection{Validation}

The validation step ensures that values previously read by the client remain
unchanged until the transaction commits.  To do this, servers ensure that the
value the client read during its transaction is the same value currently in the
data store, and that no concurrent transactions change the value that the client
read.  Specifically, servers check each transaction to ensure that it does not
read values written by, or write values read by, previously validated
transactions.  Servers also check each value against the latest value in their
local store to ensure that the value was not changed by a previously-committed
transaction.  Thus, acyclic transactions employ optimistic concurrency
control~\cite{DBLP:journals/tods/KungR81,bocc}.

Servers perform validation for each transaction before forwarding it to
subsequent servers in the chain.  This ensures that at any step in a
transactions' processing, a prefix of the chain guarantees that the transaction
is valid and will remain valid until the transaction commits or aborts.  Storage
servers will abort subsequent transactions whose writes would invalidate
previously valid transactions.  Consequently, when a transaction reaches the
last server in the chain, that server can decide to commit or abort the
transaction without consulting any other server---the protocol guarantees to the
server that every previous server will be able to commit the transaction.

When a server determines that a transaction does not validate, the server aborts
the transaction by sending an abort message backwards through the chain.  Each
server in the prefix aborts the transaction and forwards the abort message until
the message reaches the client.  These servers remove the transaction from their
local state, enabling other transactions to validate in its place.

\subsubsection{Ordering}

The central task of the acyclic transactions protocol is to establish a
serializable order across all validated transactions.  While the protocol could
simply serialize all transactions---which would maximize spurious
coordination---it instead relies upon the observation that a set of transactions
are serializable if the dependency graph of their relative orders is free of
cycles.  Each edge in this graph specifies a pair of transactions and the
relative order between them.  We refer to the transactions at each endpoint of
an edge as a {\em conflicting pair}, because one transaction contains a write
operation on a key which was read or written by the other.  Consequently, every
conflicting pair has at least one, and possibly several, keys that are common to
both transactions.

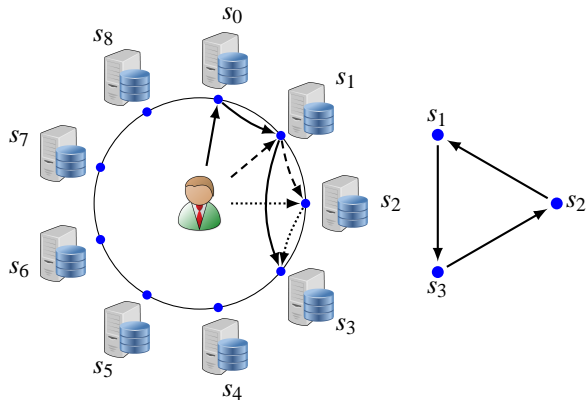
\begin{figure}[t]
\begin{center}
\begin{tikzpicture}
\usetikzlibrary{arrows}
\usetikzlibrary{calc}
\usetikzlibrary{decorations.pathreplacing}
\usetikzlibrary{shapes}

\pgfdeclareimage[height=0.75cm]{dbserv}{figs/database_server.ps};
\settoheight{\imageheight}{\pgfuseimage{dbserv}};
\pgfdeclareimage[height=0.75cm]{client}{figs/business_person.ps};
\settoheight{\imageheight}{\pgfuseimage{client}};


\draw (0em, 0em) circle (4em);

\foreach \name/\angle in {DBServ1/0, DBServ2/40, DBServ3/80, DBServ4/120, DBServ5/160,
                          DBServ6/200, DBServ7/240, DBServ8/280, DBServ9/320}
    \node[circle,color=white,fill=blue,inner sep=0pt,minimum width=0.35em] (dot\name) at (\angle:4em) {};

\foreach \name/\angle in {DBServ1/0, DBServ2/40, DBServ3/80, DBServ4/120, DBServ5/160,
                          DBServ6/200, DBServ7/240, DBServ8/280, DBServ9/320}
    \node[align=center] (\name) at (\angle:5.5em) {\pgfuseimage{dbserv}};

\foreach \name/\angle/\lbl in {DBServ1lbl/0/$s_2$, DBServ2lbl/40/$s_1$,
                               DBServ3lbl/80/$s_0$, DBServ4lbl/120/$s_8$,
                               DBServ5lbl/160/$s_7$, DBServ6lbl/200/$s_6$,
                               DBServ7lbl/240/$s_5$, DBServ8lbl/280/$s_4$,
                               DBServ9lbl/320/$s_3$}
    \node[align=center] (\name) at (\angle:7.25em) {\lbl};


\node[align=center] (client) at (0em, 0em) {\pgfuseimage{client}};

\draw[->,>=latex,thick,solid] (client) -- (dotDBServ3);
\draw[->,>=latex,thick,solid] (dotDBServ3) edge [bend right=10] (dotDBServ2);
\draw[->,>=latex,thick,solid] (dotDBServ2) edge [bend right=20] (dotDBServ9);

\draw[->,>=latex,thick,densely dashed] (client) -- (dotDBServ2);
\draw[->,>=latex,thick,densely dashed] (dotDBServ2) edge [bend right=10] (dotDBServ1);

\draw[->,>=latex,thick,densely dotted] (client) -- (dotDBServ1);
\draw[->,>=latex,thick,densely dotted] (dotDBServ1) edge [bend right=10] (dotDBServ9);

\node[right of=client,node distance=10.5em] (anchor) {};

\foreach \name/\angle in {Node1/0, Node2/120, Node3/240}
    \node[circle,color=white,fill=white,inner sep=0pt,minimum width=0.75em] (\name) at ($(anchor) + (\angle:3em)$) {};

\foreach \name/\angle in {Node1/0, Node2/120, Node3/240}
    \node[circle,color=white,fill=blue,inner sep=0pt,minimum width=0.45em] (b\name) at ($(anchor) + (\angle:3em)$) {};

\draw[->,>=latex,thick,solid] (Node1) -- (Node2);
\draw[->,>=latex,thick,solid] (Node2) -- (Node3);
\draw[->,>=latex,thick,solid] (Node3) -- (Node1);

\node[right,align=left] at (bNode1) {$s_2$};
\node[above,align=center] at (bNode2) {$s_1$};
\node[below,align=center] at (bNode3) {$s_3$};

\end{tikzpicture}
\end{center}
\caption{\small Ordering transactions in a serializable fashion across multiple
    servers is difficult because of the possibility of distributed cycles.  In
    this figure, the three transaction's chains overlap on servers $s_1$, $s_2$,
    $s_3$, but that no one server handles all three transactions.  The acyclic
    transactions protocol prevents these transactions from forming the cycle
    shown on the right.}
\label{fig:cycles}
\end{figure}

The difficulty in ordering these conflicting pairs lies not in resolving
pairwise relationships, but in ensuring that every pairwise ordering is
consistent with the globally serializable order.  Resolving the order across
multiple pairs is a considerably more complex task, because interactions between
transactions can span multiple servers.  In these cases, it is possible that no
single server would have have the requisite view to detect and prevent a cycle
in the graph.  For example, imagine three transactions across keys $k_A$, $k_B$,
and $k_C$, where each transaction writes to a different pair of the keys: $(k_A,
k_B)$, $(k_B, k_C)$, and $(k_C, k_A)$.  If the system only ordered the
transactions pairwise, the three transactions could be committed in a
non-serializable order, because no single key is written by all three
transactions.  Figure~\ref{fig:cycles} depicts this example and highlights the
problematic execution that results in a cycle between the transactions.

Servers ensure that all transactions commit in a serializable order across
servers by embedding ordering information, called {\em mediator tokens}, into
transactions.  Mediator tokens are integer values that are assigned to
transactions by the heads of chains.  Because mediator tokens are integers,
servers may determine the relative order of conflicting pair by comparing their
mediator tokens.  A simple invariant that ensures serializability is to commit
conflicting pairs in the order specified by their mediator tokens.  For example,
if the mediator tokens for the conflicting pair $(T_X, T_Y)$ have the
relationship $\mediator(T_X) < \mediator(T_Y)$, then all servers order the
transactions such that $T_X$ commits before $T_Y$.

The acyclic transactions protocol relies on this invariant to order
transactions.  Upon receipt of a transaction passing forward through the chain,
a server compares the transaction's mediator token to the largest mediator token
across all transactions that previously read or wrote any of the current
transaction's objects.  If the current mediator token is larger than the
previous token, the transaction is forwarded to the next server in the chain.
If, however, the mediator token is less than the previous token, a ``retry''
message is sent backwards in the chain to the head, where the transaction will
be retried with a larger mediator token.

\subsubsection{Generating Mediator Tokens}

At first glance, mediator tokens may resemble timestamps that are typically used
by transaction commit protocols, but mediator tokens are significantly more
flexible.  Systems based upon timestamps, whether they be logical
timestamps~\cite{!author=lamport:title=ordering:year=1978} or synchronized wall
clocks, impose restrictions on how timestamps may be generated.  Namely,
timestamps must be generated in a monotonic fashion so that the system never
moves backward in time.  Failing to preserve the monotonicity of timestamps
would permit transactions to commit in an unserializable fashion.  Mediator
tokens impose no such restrictions on token generation.

The flexibility of mediator tokens permits a wide array of implementation
strategies.  For example, a simple token generation strategy would be to always
set a transaction's initial token to zero, choosing successively larger values
on each subsequent pass.  Another strategy would be to generate tokens at random
and ensure that each subsequent pass draws from a range of tokens that are
strictly greater than the token of the previous pass.  A strategy that limits
the number of retries is for each server to maintain a counter to generate
mediator tokens.  Servers may generate a new mediator token by reading the
counter's current value and incrementing the counter.  When a server sees a
mediator token that is greater than the next value to be generated by its
counter, it may advance the counter to be greater than this other token.
Because mediator tokens are flexible, storage servers do not need to carefully
manage or preserve this counter during server failover, and do not need to
synchronize counters across servers.

\subsection{Fault Tolerance and Durability}
\label{sec:fault-tolerance}

In a large-scale deployment, failures are inevitable.  Acyclic transactions
accommodate a natural way to overcome such failures.  Specifically, acyclic
transactions permit a subchain of $f + 1$ replicas to be inlined into the longer
chain in place of a single data server.  This allows the system to remain
available despite up to $f$ failures within a subchain.  Chain replication
maintains a well-ordered series of updates within each subchain.  Operations
that traverse the acyclic transaction chain in the forward direction pass
forward through all inlined chains.  Likewise, operations that traverse the
chain in reverse traverse inlined chains in reverse.

The notion of fault-tolerance provided by acyclic transactions is different from
the notion of durability within traditional databases.  While durability ensures
that data may be re-read from disk after a failure, the system remains
unavailable during the failure and recovery period; in contrast, acyclic
transactions' fault tolerance mechanism ensures that the system remains
available so long as the number of failures remains below the configured
threshold.

\subsection{Atomicity, Consistency, Isolation}
\label{sec:aci}

The protocol guarantees atomicity, consistency, and isolation for all
transactions.  These properties naturally follow from the one-copy
serializability upheld by the protocol.  Each transaction completes in its
entirety at a well-defined point in the partial order, where its effects are
either completely visible to subsequent transactions, or it aborts without
effect.  Every server ensures that the stored objects are well-formed and match
their data types.  Overall, acyclic transactions guarantee that operations
within a transaction execute with mutual exclusion from each other, as if there
were a single giant lock protecting the database.

\subsection{Correctness}

By leveraging a fault tolerant system coordinator, acyclic transactions uphold
both liveness and safety in the presence of up to $f$ faults.  Specifically,
acyclic transactions maintain serializability at all times, and will eventually
commit or abort every transaction assuming at most $f$ of the $f + 1$ replicas
of the data remain non-faulty.  In this section we demonstrate how the acyclic
transactions protocol maintains these safety and liveness properties.

\paragraph{Safety:}  The acyclic transactions protocol provides serializability
by ensuring that the final system state is equivalent to a serial schedule.  The
protocol upholds this guarantee by ensuring that the dependency graph across all
transactions is acyclic.  Intuitively, every conflicting pair directly
corresponds to an edge in this graph, while the mediator tokens enforce the
anti-cycle property.

Because non-conflicting pairs operate on disjoint sets of data, the conflicting
pairs are the only transaction pairs whose order must be carefully managed.  A
conflicting pair of transactions is committed in the same order on all common
servers in order to ensure that operations to their shared state are applied in
the same order.  Servers use mediator tokens to decide the commit order for
transactions in a conflicting pair; every server will enforce the order
specified by the transactions' tokens.

Globally, mediator tokens preserve transitive relationships across transactions,
ensuring that cycles cannot arise in the dependency graph.  The dependency graph
is structured such that each edge is a conflicting pair $(T_x, T_y)$ such that
either $\mediator(T_x) < \mediator(T_y)$ or $\mediator(T_x) > \mediator(T_y)$.
In the former case, the graph will contain an edge $T_x \leadsto T_y$, while in
the latter case, the graph will contain edge $T_y \leadsto T_x$.  Any directed
path will consist of edges such that the mediator token for the source is less
than the mediator token for the destination.  Transitively, the start of the
path must have a mediator token less than the end of the path.  Thus, it is
impossible for the graph to contain a cycle, because a cycle would imply that
there exists a directed path---and thus, a directed edge---from a transaction
with a higher mediator token to a transaction with a lower mediator token.
Because the system prohibits committing transactions in an order that
contradicts the ordering established by their mediator tokens, it is impossible
for such an edge, and thus a cycle, to exist.

\paragraph{Liveness:}  The protocol remains available for processing
transactions during a bounded number of server failures.  Specifically, the
protocol will always be able to commit or abort a transaction so long as at most
$f$ servers fail of the $f+1$ servers assigned to replicate each key.  To enable
the system to detect and correct for these failures, acyclic transactions make
use of a fault tolerant coordinator, which may be built using standard
techniques~\cite{!burrows:chubby,zookeeper,openreplica}.  This coordinator acts
as a shepherd for the system, guiding it toward a stable state, even as servers
fail or become otherwise unavailable.

The system overcomes failures by removing failed servers from the chains for
actively propagating transactions.  For each failure, the coordinator issues a
new configuration that lists the server as failed.  Non-faulty servers may
consult this new configuration to determine which currently outstanding
transactions contain the faulty server as the next hop in the forward direction.
These transactions are retransmitted to move the transaction toward a commit or
abort state.  To prevent duplicate messages from affecting correctness, servers
maintain a list of committed and aborted transactions.  Upon receipt of a
retransmitted forward-bound message, a server will first consult this list and
answer with a commit or abort message if appropriate.  Otherwise, the server
processes the message to move the transaction closer to committing; typically
this will entail sending another message forward in the chain, or waiting for a
previously sent message to return ``commit'' or ``abort''.  Overall, the
coordinator and servers will repeat this process as necessary until transactions
eventually commit or abort.

\section{Implementation}
\label{sec:implementation}

We have fully implemented the system described in this paper.  The code base
consists of ~130,000 lines of code, approximately 15,000 lines of which are
devoted to processing transactions.  The Warp distribution provides bindings for
C, C++, Python, Ruby, Java, Go, and Node.JS and supports a rich API that goes
well beyond the simple \texttt{get}/\texttt{put} interface of typical key-value
stores.  A system of virtual servers maps a small number of servers to a larger
number of partitions, permitting the system to reassign partitions to servers
without repartitioning the data.  The implementation uses a replicated state
machine as the coordinator to ensure that there are no single points of failure.

\subsection{Rich API}
\label{sec:rich-api}

Acyclic transactions naturally support an expanded API that enables complex
applications.  The expanded API includes support for rich data structures,
multiple independent schemas, and nested transactions.

\subsubsection{Data Structures}

The discussion in Section~\ref{sec:arch} presented all operations in a acyclic
transaction as either a read or a write on an arbitrary string, but our
implementation goes much further to support many data structures commonly used
in modern applications.  Warp provides programmers with integer, float, list,
set, map, and document types as well as atomic operations on each of these types
that enable fast concurrent operation.  For example, it is possible to
atomically add an element to a list, or perform arithmetic on an integer type.
A write, then, may consist of any of these atomic operations and is not limited
to simply overwriting the previous value.  These atomic operations are
especially useful for cases where acyclic transactions enable low abort and
retry rates because they allow applications to further improve concurrency.

\subsubsection{Independent Schemas}

Acyclic transactions generalize well from operations across multiple keys to
operations across multiple keys in different schemas.  In our implementation,
applications may create multiple schemas---which resemble tables from
traditional database systems---and store different objects in each schema
without any collisions in the key space.  Clients construct the chain for
transactions that touch multiple schemas by lexicographically ordering servers
first by schema, then by key.

\subsubsection{Nested Transactions}

The architecture we have presented naturally supports nested transactions with
only minimal changes to the client library.  Nested transactions may be
implemented by allowing transaction contexts to recursively refer to each other.
Each nested transaction maintains its own locally-managed transaction context
with a pointer to the parent transaction's context.  Reads recursively query the
parent context until either a cached value is read, or the root context issues
the query to a storage server.  Writes are stored in the transaction context to
which they are issued.  At commit time, the client merges a nested transaction
into its parent context, by merging the read and write sets.  Nested
transactions abort if the values read in the child are modified in the parent or
vice-versa.  The client sends a acyclic transaction to the storage servers only
when the root transaction commits.

\subsection{Virtual Servers}
\label{sec:virtual}

Warp uses a system of virtual servers to map multiple partitions of the mapping
to a single server.  Clients construct their acyclic transaction chains by
constructing a chain through the virtual servers, and then mapping these virtual
servers to their respective servers.  A server that maps to multiple virtual
servers in a chain will appear at multiple places in the chain, where it acts as
each of its virtual servers independently.  Within each physical server, state
is partitioned by virtual server, so that each virtual server functions as if it
were independent.  Virtual servers enable the system to perform dynamic load
balancing more efficiently.

\subsection{Coordinator}
\label{sec:coordinator}

A replicated state machine called the coordinator partitions the key space
across all data servers, ensures balanced key distribution, and facilitates
membership changes as servers leave and join the cluster.  Since the coordinator
is not on the data path, its implementation is not critical to the performance
of acyclic transactions.

The coordinator partitions data across servers and ensures balanced key
distribution by using copyset replication~\cite{copysets} to group servers into
replica sets.  Each independent schema is partitioned across the generated
copysets to create an object-to-server mapping.  The coordinator over-partitions
the key space to enable it to remap partitions from over-burdened replica sets
to under-loaded replica sets if necessary.

As servers join and leave a cluster, the coordinator regenerates copysets to
respond to new members.  Servers dynamically compute the previous and next
servers in each acyclic transaction's chain using the mapping; when the mapping
changes, servers retransmit transactions whose chain changed.  Every message
carries the configuration's version to enable clients and servers to detect and
re-route out-of-date requests using an up-to-date configuration. The mapping is
changed incrementally, ensuring that each subsequent mapping overlaps with the
previous mapping, which ensures that some replicas in each inlined chain will
overlap as well.  Thus, servers are always able to integrate new nodes without
violating the assumptions used to construct acyclic transactions' chains.

The coordinator is implemented on top of the Replicant replicated state machine
system.  Replicant uses chain replication~\cite{!schneider:chain} to sequence
the input to the state machine and a quorum-based protocol to reconfigure chains
on failure.  The details of Replicant are beyond the scope of this paper; the
function of the coordinator could also be built on configuration services such
as Chubby~\cite{!burrows:chubby}, ZooKeeper~\cite{zookeeper}, and
OpenReplica~\cite{openreplica}.

\section{Evaluation}
\label{sec:evaluation}

In this section, we evaluate Warp's performance and scalability using both macro
and micro benchmarks.  The primary focus of our evaluation is on examining the
performance of Warp transactions relative to other transaction processing
techniques.  To that end, we implemented Sinfonia's
mini-transactions~\cite{!aguilera:sinfonia:year=2007} on top of HyperDex,
hereafter referred to as MiniDex.  Because Warp builds upon HyperDex, and
because native HyperDex outperforms many NoSQL databases, we ensure a true
apples-to-apples comparison by building all systems using the same code base.
We also compare Warp to HyperDex, even though the latter offers no transactional
guarantees.  The client-facing interfaces and the benchmark code is identical
for all three systems.

We performed our experiments on our dedicated lab-size cluster consisting of
thirteen servers, each of which is equipped with two Intel Xeon
\unit{2.5}{\giga\hertz} E5420 processors, \unit{16}{\giga\byte} of RAM,
\unit{500}{\giga\byte} SATA \unit{3.0}{\giga\bit\per\second} hard disks, and
Gigabit Ethernet.  The servers are running 64-bit Ubuntu 14.04.  Each storage
system was configured with appropriate settings for a real deployment of this
size. This includes setting the replication factor to be the minimum value
necessary to tolerate one failure of any process or machine.  Both the
coordinators and the storage servers can each tolerate one failure.  All systems
provide strong consistency guarantees, which MiniDex and Warp extend across
multiple objects.

\subsection{TPC-C Macro Benchmark}

\begin{figure}[t]
\flushright
\input{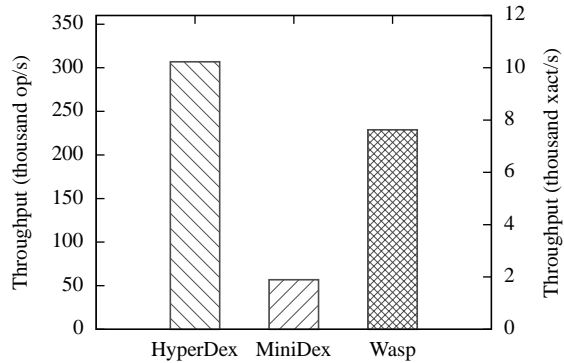}
\caption{\small Total transactional throughput of the three systems.  Warp
    outperforms MiniDex by a factor of 4, and achieves 75\% the throughput of
    HyperDex, which Warp uses as its underlying key-value store.  Warp averages
    approximately 7,500 transactions, or more than 225,000 individual key
    operations, per second in this benchmark.}
\label{fig:tpcc:throughput}
\end{figure}

\begin{figure*}[t]
\begin{subfigure}[b]{0.45\textwidth}
\centering
\input{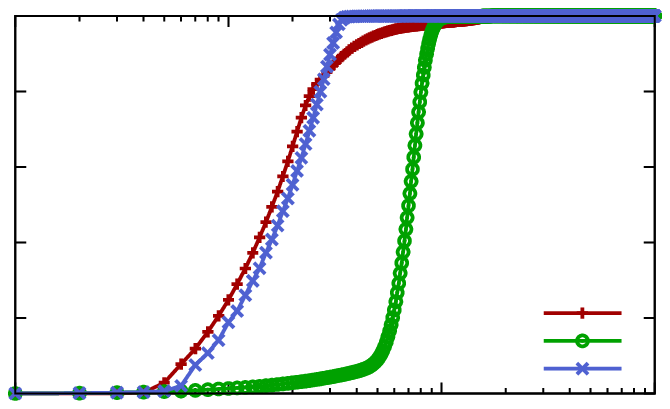}
\caption{\small New Order Transactions}
\label{fig:tpcc:latency:new-order}
\end{subfigure}%
\qquad%
\begin{subfigure}[b]{0.45\textwidth}
\centering
\input{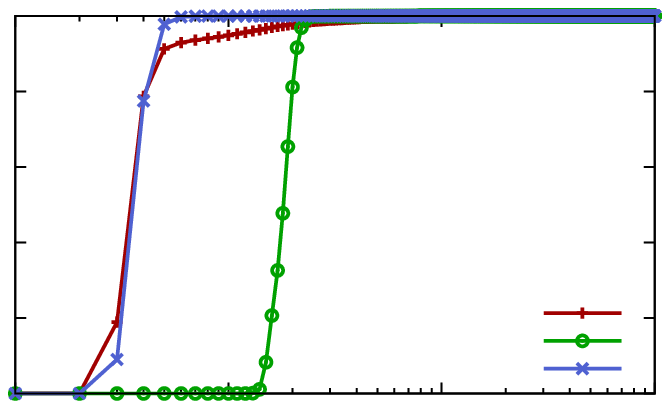}
\caption{\small Payment Transactions}
\label{fig:tpcc:latency:payment}
\end{subfigure}
\caption{\small Write operations have similar latency in HyperDex and Warp
    because the cumulative lengths across all chains formed within a transaction
    are the same.  A HyperDex transaction with $O$ operations makes $O$ chains
    of length $f + 1$.  A Warp transaction will make a single chain of length $O
    \times (f + 1)$.}
\label{fig:tpcc:latency}
\end{figure*}

The industry-standard TPC-C benchmark simulates an e-commerce application by
specifying a mixed transaction workload.  The workload specified by TPC-C is
inherently difficult to process with optimistic concurrency control, because it
includes both read-heavy and update-heavy transaction profiles and the
update-heavy transactions intentionally contend on a small number of hot keys.
For instance, the {\em new-order} transaction generates the order's identifier
using a sequentially-increasing counter associated with one of one-hundred
districts.  The {\em payment} transaction increments the year-to-date totals for
one of one-hundred districts and one of ten warehouses.  The contention and
interaction between the new-order and payment transaction profiles is what makes
the TPC-C benchmark a compelling choice for testing new optimistic protocols.

\begin{table}
\centering
\begin{tabular}{lcccc}
\hline
Profile     & R     & W     & RMW   & \% Mix \\
\hline
New Order   & 12    & 3     & 11 (1)& 45 \\
Payment     & 0     & 1     & 3 (2) & 45 \\
Order Status & 12   & 0     & 0     & 5 \\
Stock Level & 201 (1) & 0   & 0     & 5 \\
\hline
\end{tabular}
\caption{\small A summary TPC-C workload.  For each transaction profile, the
    chart shows the average number of read-only (R), write-only (W), and
    read-modify-write (RMW) operations.  The TPC-C workload will randomly
    perform transactions accordingo to this distribution.
The values in parenthesis specify the numberj
of district or warehouse objects per transaction.}
\label{tab:tpcc:overview}
\end{table}

At the core of the benchmark are multiple transaction profiles which each
represent a different type of application logic.  Table~\ref{tab:tpcc:overview}
provides an overview of each transaction type.  The values in parenthesis
specify the number of district or warehouse objects per transaction.  The bulk
of the workload stems from the new-order and payment transaction profiles.
These profiles simulate a customer placing purchase orders, and subsequently
paying the invoice.  Our implementation of TPC-C retains as much functionality
of the benchmark as is reasonable to implement on a key-value store.  In total,
the implementation consists of approximately 1100 lines of Python code that
execute client side using the Python bindings to HyperDex, MiniDex, and Warp.
We omitted the ``delivery transaction'' profile because the TPC-C benchmark
specifies that it be performed by a background process that would be handled by
a messaging queue in a real deployment.  Because we chose to retain most of the
TPC-C benchmark's behavior, our results are incomparable to others in the
literature that simply perform new-order
transactions~\cite{!abadi:calvin,iconf}.

We deployed the TPC-C benchmark with its default setting that includes 10
warehouses, which are very contended keys, and 100 districts, which are somewhat
contended keys.  Each new-order or payment transaction includes one warehouse
and one district in the set of keys that it reads, modifies, and writes.  For
the transactional systems, these keys will be the ones most likely to introduce
transaction abort and retries.  For the HyperDex workload, there cannot possibly
be any conflict because the reads and writes may proceed in any order without
transactional consistency.  Intuitively, we expect that the performance of
HyperDex provides an upper bound on throughput and a lower bound on latency for
all experiments, because MiniDex and Warp add strictly more mechanism on top of
the existing code.  Consequently, the HyperDex upper bound allows us to
objectively gauge how much overhead each system adds to the baseline.

Figure~\ref{fig:tpcc:throughput} shows the overall transactional throughput for
HyperDex, MiniDex, and Warp.   The experiment shows that that Warp achieves a
throughput that is four times higher than MiniDex, and close to 7,500
transactions, or 225,000 operations, per second.  To put the factor of 4 in
perspective, Warp achieves 75\% the throughput of the non-transactional system
on which it builds, while MiniDex does not even realize 20\% of its potential.

The intuition for why Warp is so much more efficient is two-fold:  first, Warp's
transaction management allows more concurrency than is possible with MiniDex;
and second, Warp's communication costs are similar to those of the baseline and
require no additional messages.  Both systems construct chains to write data
into the system, where each link in the chain equates to a network round trip.
Where HyperDex will construct one chain of length $f + 1$ for each of the $O$
operations, Warp will commit the operations through a single chain of length $O
\times (f + 1)$ to commit the transaction.  Thus, in the common case of no
aborts and no retries, Warp requires no additional round trips beyond those
required for a write within HyperDex.  Figures~\ref{fig:tpcc:latency:new-order}
and~\ref{fig:tpcc:latency:payment} show latency CDFs for the new-order and
payment transaction profiles.  We can see that for both transaction types, the
latency of HyperDex and Warp follow a similar trend, while the latency of
MiniDex is approximately five times higher.

\begin{figure}[t]
\centering
\input{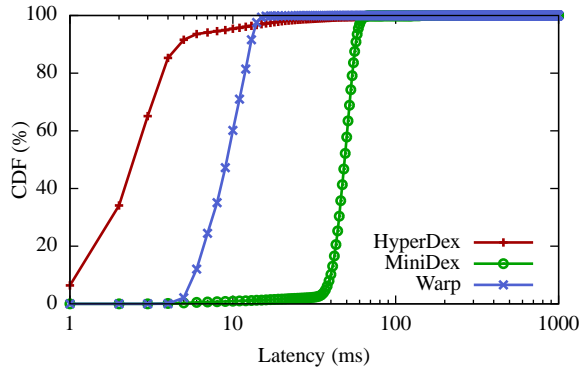}
\caption{\small Read operations have different latencies inside and outside of a
    transaction.  A non-transactional read may be directly performed against the
    server.  A transactional read includes that latency, plus the cost of
    validating the read at commit time.}
\label{fig:tpcc:latency:order-status}
\end{figure}

Because transactions must validate read operations, there's an additional cost
to performing a transactional read that is not paid for non-transactional
workloads.  The read that the Warp client library performs to pull the object
into the transaction context is the same cost as the read that the
non-transactional code will perform.  The Warp client then validates the read at
commit time.  In figure~\ref{fig:tpcc:latency:order-status}, we directly
quantify the latency profile of the read-only ``order status'' transaction.  We
can see that Warp's latency is approximately three times higher than the
non-transactional measurement, while the MiniDex latency is approximately six
times higher.

Overall, the reason MiniDex achieves lower throughput and higher latency is
because mini-transactions are more likely to abort.  We observed, on average,
only 5\% of transactions complete without aborting or retrying at least once,
and we've included the time taken to retry transactions in the above numbers for
all systems.  Because all three systems use the same benchmark and baseline
code, the performance difference is solely the commit protocol in use.  MiniDex
cannot permit multiple transactions to prepare for the same key simultaneously,
forcing transactions to abort or wait, which increases the latency by a small
constant multiplier.  Warp permits these transactions to prepare simultaneously,
enabling it to complete all transactions without aborting.

Although it may seem possible to relax the mini-transactions protocol to permit
transactions to prepare for the same key simultaneously, doing so would break
serializability.  A modified MiniDex would require additional mechanisms to
prevent the potential cycle illustrated in Figure~\ref{fig:cycles}, as
concurrently prepared transactions could commit in different orders on different
servers.  Even HyperDex's atomic operations cannot enable such a relaxed commit
protocol because they cannot affect the order in which operations occur on
different servers.

\subsection{Micro-Benchmarks}

In order to gain insight into the behavior of Warp's acyclic transactions, we
examine the results from several targeted micro-benchmarks.  In all of these
micro-benchmarks, objects have \unit{12}{\byte} keys and \unit{64}{\byte}
values, and are constructed uniformly at random.  Ten million objects are
preloaded onto the cluster before performing each benchmark.

\subsubsection{Read/Write Ratio}

\begin{figure}[t]
\flushright
\input{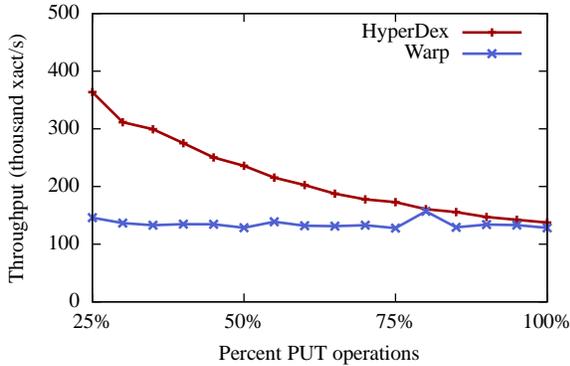}
\caption{\small The ratio of read/write operations does not materially affect
    the throughput for transactions.}
\label{fig:ratio}
\end{figure}

In order to quantify the effects of the read/write ratio on a transactions'
throughput, we constructed a micro-benchmark that varies the read-write ratio
for operations of constant size.  This micro-benchmark constructs transactions
that involve exactly eight objects, and randomly read from or write to random
objects.  Each operation is randomly chosen to be a read or a write so that the
total percentage of write operations matches the independent variable.  In
HyperDex, a read incurs one round trip, while a write incurs $f + 1$ round
trips.  Thus, we expect that a write-heavy workload in HyperDex will achieve
lower throughput than a read-heavy workload, with all other factors fixed.
Because of the validation step, we expect Warp transactions to be largely a
matter of the latency of the commit.  In Figure~\ref{fig:ratio} we see the
average throughput for Warp is 150,000 transactions per second, regardless of
the workload, while HyperDex performance increases as read percentage increases.
It demonstrates that the performance is a function of the transaction protocol
and not the read-write ratio.

\subsubsection{Transaction Size}

\begin{figure}[t]
\begin{flushright}
\input{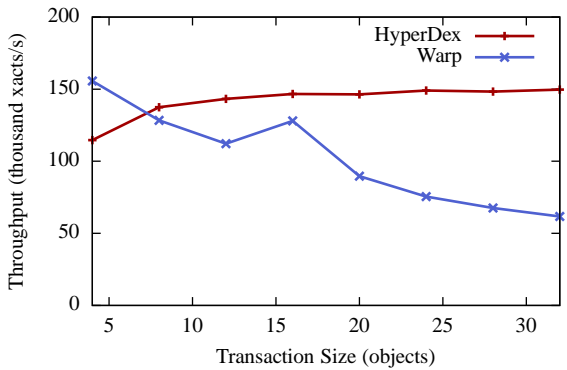}
\caption{\small The total throughput of Warp is dependent on the throughput of
    the underlying key-value store, and of the transaction size.  This graph
    shows the throughput of a 100\% write workload as the number of keys in a
    transaction increases.}
\label{fig:batch}
\end{flushright}
\end{figure}

Naturally, the use of chains introduces a tradeoff:  as transactions grow to
contain more keys, the length of the resulting chains naturally increases as
well.  Figure~\ref{fig:batch} quantifies this tradeoff by constructing write
transactions with different numbers of keys.  We employ the same micro-benchmark
from the previous section, and use a 100\% write workload.

To test the performance impact of transaction size, we modified our previous
microbenchmark to vary the number of keys in a transaction rather than the
read/write ratio. In this experiment, the microbenchmark issues transactions
with a configurable number of \texttt{put} operations on random keys.
Figure~\ref{fig:batch} shows that, as expected, the number of operations per
second is mostly independent of the transaction size. This demonstrates that
longer transaction chains do not introduce additional overhead, and that, for
this workload, the transaction rate is a linear function of the transaction
size.

\subsubsection{Scalability}

\begin{figure}[t]
\begin{flushright}
\input{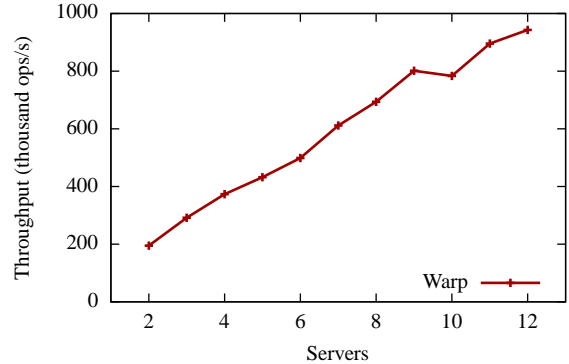}
\caption{\small Warp is a scalable system.  This graph shows the aggregate throughput
    of the system as servers are added.  With each additional server, the
    overall throughput increases proportionally, exhibiting linear scaling.}
\label{fig:scalability}
\end{flushright}
\end{figure}

The performance of acyclic transactions should scale linearly with the number of
servers in the cluster, as the number of servers that participate in a acyclic
transaction is dependent only on the transaction size.  Adding more servers to
the cluster should therefore yield a proportional increase in performance by
spreading the work across more servers.  Figure~\ref{fig:scalability} shows the
aggregate throughput of a two-key transaction from our micro-benchmark with
different cluster sizes.  Not surprisingly, Warp throughput scales linearly with
cluster size.

\subsection{Summary}

Overall, the acyclic transactions protocol provides a low overhead means of
ensuring serializable transactions in a distributed key value store.  Despite
providing ACID guarantees, Warp achieves throughputs close to those associated
with non-transactional data stores.  This is primarily due to the way acyclic
transactions constrict data to chains that avoid bottlenecks at dedicated
transaction managers.  This level of performance does not require static
analysis of applications, or dependency on synchronicity assumptions, or
reliance on loosening application semantics.

\section{Related Work}
\label{sec:background}

Transaction management has been an active research topic since the early days of
distributed database systems.  Existing approaches can be broadly classified
into the following categories based upon the mechanisms employed and resulting
guarantees.

\paragraph{Optimistic Concurrency Control:}
Acyclic transactions are a form of optimistic concurrency
control~\cite{DBLP:journals/tods/KungR81} because a validation step is necessary
to prevent conflicts among concurrent transactions.  Traditionally, optimistic
concurrency control schemes have been divided into backward-oriented and
forward-oriented schemes~\cite{bocc}.  The former checks optimistic reads
against previously performed writes, while the latter checks optimistic writes
against concurrently executing, unvalidated reads.  Warp's concurrency control
is most similar to backward oriented OCC, but includes additional validation
steps beyond those typically employed in backward oriented systems because it
permits transactions to execute concurrently.

\paragraph{Centralized:}
Early RDBMS systems relied on physically centralized transaction
managers~\cite{DBLP:conf/vldb/ChamberlinGY81}.  While centralization greatly
simplifies the implementation of a transaction manager, it poses a performance
and scalability bottleneck and is a single point of failure.  Warp is based on a
distributed architecture.

\paragraph{Distributed:}
The traditional approach to distributing transaction management is to provide a
set of specialized transaction managers that serve as intermediaries between
clients and back-end data servers.  These transaction managers perform lock or
timestamp management~\cite{!bernstein-goodman:concurrency:year=1981}, and employ
a protocol, such as two phase-commit, for coordination.  In the class of
two-phase commit protocols, Linear 2PC~\cite{linear2pc} is similar to Warp in
that the communication pattern is arranged along a chain of servers.  The key
improvement of acyclic transactions over existing multi-phase protocols,
including Linear 2PC, is that it does not employ specialized transaction
managers and permits multiple transactions to execute in parallel directly on a
subset of storage servers, even when operating on the same key(s).

A recent proposal~\cite{!lomet:unbundling} suggests physically separating the
transaction processing component from the storage component so that transaction
processing remains agnostic to the structure of the storage.  The resulting
system, Deuteronomy~\cite{!lomet:deuteronomy} performs this separation so that
transaction management remains isolated from scaling decisions made at the
storage layer.  ElasTraS~~\cite{!author=das:title=elastras:year=2013} uses two
layers of transaction management to separately process read-only and read-write
transactions, where each layer and the underlying storage are independently
scalable.  Separating transaction management from data storage does not
fundamentally make the transaction management more scalable because it does not
alter the spurious coordination naturally present in the transaction manager.
Warp reduces spurious coordination and centralized bottlenecks by employing a
completely distributed protocol.

Recent work has proposed database systems with minimal coordination through the
use of a technique called $\mathcal{I}$-confluence analysis~\cite{iconf}.
$\mathcal{I}$-confluence requires of the programmer a set of invariants that are
processed by offline analysis to determine a minimal amount of coordination to
uphold the invariants.  Warp transactions are fully serializable, require no
programmer-provided specifications, and naturally avoid spurious coordination.

\paragraph{Consensus-based:}
Recent work has examined how to use a general consensus protocol, such as
Paxos~\cite{!lamport:parliament} or Zab~\cite{zookeeper}, to serialize
transactions in a fault-tolerant manner.  Although consensus seems unrelated to
transaction management, the classic two-phase commit algorithm is actually a
special $f=0$ case of Paxos that cannot tolerate coordinator
failure~\cite{!lamport:transaction}.

Straightforward application of consensus protocols, however, would introduce
spurious coordination by applying a total order across all transactions.
Consequently, consensus-based systems typically use some combination of data
partitioning~\cite{scalaris,!arvind:scatter,!corbett:megastore,sdur},
Generalized Paxos~\cite{!madden:mdcc} or transaction
batching~\cite{!abadi:calvin,sdur-geo} to increase opportunities for parallel
execution.

Warp uses consensus only to maintain system meta-state.  The acyclic
transactions protocol, inspired by chain-replication~\cite{!schneider:chain} and
value-dependent chaining~\cite{!escriva:hyperdex}, relies upon consensus for
system membership and coordination, but not for actual transaction processing.
Because consensus is not on the critical path for any acyclic transaction, the
protocol is able to completely eliminate any consensus-induced overhead.

\paragraph{Synchronized clocks:}
Some notable systems in this space take advantage of synchronized clocks to
order transactions.  Adya et.~al.~\cite{!adya:optimistic-synchronized} support
serializable transactions and use loosely synchronized clocks as a performance
optimization.  Spanner~\cite{!author=hsieh:title=spanner:booktitle=osdi} uses
tightly synchronized clocks, with bounded error, to achieve high-throughput and
external consistency for transactions across multiple data centers.
Granola~\cite{!liskov:granola} orders independent transactions with no locking
overhead or abort mechanism, and orders these transactions using time
synchronization as an optimization.

Warp is an asynchronous protocol that makes no assumptions about clock
synchrony.  It is more robust than systems which make synchronicity assumptions,
and requires less maintenance and operations infrastructure.  Most systems in
this category remain correct should synchronicity assumptions be violated, but
suffer varying degrees of performance degradation.  A notable exception is
Spanner, which preserves serializability only when its assumptions are upheld.

\paragraph{Client-managed transactions:}
Some systems build on existing storage by implementing transactions directly in
the client library.  Such systems mediate concurrent transactions by embedding
additional attributes into the stored objects to enable concurrency control.
CrSO~\cite{!flavio:transactional} uses HBase versions and a centralized status
oracle to check for read-write or write-write conflicts at commit time.
Percolator~\cite{!dabek:distributed-transactions} maintains Google's search
index by storing both data and locks in BigTable.

The downside to client-managed approaches is that they require mechanisms to
cope with client failure.  CrSO requires a background process to cleanup stale
versions of objects written by failed transactions.  Percolator uses a
background mechanism to break locks held by failed processes.  Warp incurs no
such cost because failed clients leave behind no state to clean up.

\paragraph{Geo-Replication:}
For geo-replicated storage, many systems avoid synchronous WAN latencies by
making guarantees weaker than serializability.  COPS-GT~\cite{!lloyd:cops} and
Eiger~\cite{!lloyd:stronger} provide read and write transactions, respectively,
that commit locally and propagate to remote data centers in a
causally-consistent fashion.  Walter~\cite{!sovran:transactional} implements
parallel snapshot isolation using counting sets to resolve conflicting versions,
similar to commutative data types~\cite{!shapiro:crdts}.  Warp provides a
strictly stronger guarantee of general purpose serializable transactions, but
lacks optimizations for geo-replication.

\paragraph{Offline Checking:}
Lynx~\cite{lynx} uses chains for replication and guarantees serializability, but
requires a priori knowledge of transactions and static analysis to prevent
non-serializable executions.  The key insight in Lynx is that this static
analysis can break one application-level transaction into many smaller
transactions that execute piecewise across servers.  Rococo~\cite{rococo} also
requires offline analysis of transactions in order to decompose them into
smaller atomic units, which it then executes in parallel.  Warp is fundamentally
different from transaction management schemes that rely upon offline checking
because it guarantees serializability without requiring that transactions are
known a priori, and without requiring any static analysis across all
transactions.

\paragraph{Workload Partitioning:}
Some systems improve performance by constraining transactions to operate within
single partitions of the data store.  G-Store~\cite{!agrawal:g-store} provides
serializable transactions on top of HBase by grouping keys' primary replicas on
a single server so that transactions require no cross-server communication.
H-Store~\cite{!stonebraker:complete-rewrite} targets OLTP applications and
efficiently supports such constrained tree applications by guaranteeing that
transactions are executed by a single server.  Warp imposes no constraint on
transactions, enabling maximal flexibility in data placement.

\paragraph{Mini-Transactions:}
Sinfonia~\cite{!aguilera:sinfonia:year=2007} introduces the mini-transaction
primitive which allows an application to specify sets of checks, reads, and
writes and commit the result using a modified two-phase commit.  The payload of
a mini-transaction is the same as the payload of a acyclic transaction.  Indeed,
the protocols can be quite similar in their behavior: both are optimistic, both
require two (or more) phases to commit a transaction across multiple servers,
and both have the client library optimistically execute reads and writes to be
validated at commit time.

Where mini-transactions and acyclic transactions differ is in their commit
behavior.  Acyclic transactions allow multiple transactions that read or write
the same key to simultaneously execute and commit in a serializable order.
Additionally, mini-transactions are vulnerable to client failure while acyclic
transactions are naturally fault tolerant.  These differences cannot be overcome
by simply loosening the commit requirements within Sinfonia, because the
resulting mechanism not be serializable.

\paragraph{Key-Value Stores:}
Key-value systems are defined by their distributed architecture that offers
performance and scalability, often obtained by avoiding strong consistency or
transactional guarantees.  This trade-off is often an engineering decision to
mask latency, and is not fundamental.  Amazon's Dynamo~\cite{!vogels:dynamo} and
its derivatives~\cite{cassandra,voldemort,riak} guarantee only eventual
consistency in order to increase write availability by writing data to sloppy
quorums.  Yahoo!'s~PNUTS~\cite{!title=pnuts:journal=pvldb} makes a slightly
stronger guarantee of per-object timeline consistency, but makes no guarantees
across multiple objects.  Google's
BigTable~\cite{DBLP:conf/osdi/ChangDGHWBCFG06} provides linearizable access to
individual rows, but does not make cross-object guarantees.  BigTable's
consistency is the same as HyperDex~\cite{!escriva:hyperdex}, the system Warp
builds upon.  Warp's guarantee is strictly stronger as it extends
serializability across multiple objects.

More generally, these NoSQL systems have roots in Distributed Data
Structures~\cite{gribble} and distributed hash
tables~\cite{!ratnasamy:content-addressable,!stoica:chord,pastry,!kubiatowicz:tapestry},
which provide efficient access to individual objects, usually in the form of a
key-value store.  Other notable work on key-value stores includes
FAWN-KV~\cite{!author=andersen:title=fawn:booktitle=sosp}, a linearizable
key-value store built to reduce power consumption in storage systems;
Comet~\cite{!author=geambasu:title=comet:booktitle=osdi}, a key-value store that
stores clients' code alongside the stored objects;
RAMCloud~\cite{!author=ousterhout:title=ramcloud:booktitle=sosp}, which builds a
key-value store for low-latency networks; and FaRM, which uses RDMA to build a
fast in-memory key-value store with single-machine transactions.  The goals of
these systems are orthogonal to those in Warp, and the techniques could be
combined to make a transactional key-value store with low power consumption
(maximizing transactions per watt), or low latency (minimizing transaction
completion time).

\section{Conclusion}
\label{sec:conclusion}

This paper describes Warp, a key-value store that provides one-copy-serializable
ACID transactions.  The main insight behind Warp is a protocol called acyclic
transactions which enables the system to completely distribute the task of
ordering transactions.  Consequently, transactions on separate servers will not
require expensive coordination and the number of servers that process a
transaction is independent of the number of servers in the system.  The system
achieves high performance on a variety of standard benchmarks, performing nearly
as well as the non-transactional key-value store that Warp builds upon.

\footnotesize
\bibliographystyle{plain}
\bibliography{warp}

\newcommand{\etalchar}[1]{$^{#1}$}
\begin{thebibliography}{00}

\bibitem{!adya:optimistic-synchronized}
Atul Adya, Robert Gruber, Barbara Liskov, and Umesh Maheshwari.
\newblock Efficient Optimistic Concurrency Control Using Loosely Synchronized Clocks.
\newblock In Proceedings of the \emph{SIGMOD International Conference on Management of Data,} pages 23-34, San Jose, California, May 1995.

\bibitem{!aguilera:sinfonia:year=2007}
Marcos Kawazoe Aguilera, Arif Merchant, Mehul A. Shah, Alistair C. Veitch, and Christos T. Karamanolis.
\newblock Sinfonia: A New Paradigm For Building Scalable Distributed Systems.
\newblock In Proceedings of the \emph{Symposium on Operating Systems Principles,} pages 159-174, Stevenson, Washington, October 2007.

\bibitem{openreplica}
Deniz Alt{\i}nb{\"u}ken and Emin G{\"u}n Sirer.
\newblock Commodifying Replicated State Machines With OpenReplica.
\newblock Cornell University, Technical Report, 2012.

\bibitem{!author=andersen:title=fawn:booktitle=sosp}
David G. Andersen, Jason Franklin, Michael Kaminsky, Amar Phanishayee, Lawrence Tan, and Vijay Vasudevan.
\newblock FAWN: A Fast Array Of Wimpy Nodes.
\newblock In Proceedings of the \emph{Symposium on Operating Systems Principles,} pages 1-14, Big Sky, Montana, October 2009.

\bibitem{iconf}
Peter Bailis, Alan Fekete, Michael J. Franklin, Ali Ghodsi, Joseph M. Hellerstein, and Ion Stoica.
\newblock Coordination-Avoiding Database Systems.
\newblock In \emph{CoRR,} abs/1402.2237, 2014.

\bibitem{!corbett:megastore}
Jason Baker, Chris Bond, James C. Corbett, J. J. Furman, Andrey Khorlin, James Larson, Jean-Michel Leon, Yawei Li, Alexander Lloyd, and Vadim Yushprakh.
\newblock Megastore: Providing Scalable, Highly Available Storage For Interactive Services.
\newblock In Proceedings of the \emph{Conference on Innovative Data Systems Research,} pages 223-234, Asilomar, California, January 2011.

\bibitem{!bernstein-goodman:concurrency:year=1981}
Philip A. Bernstein and Nathan Goodman.
\newblock Concurrency Control In Distributed Database Systems.
\newblock In \emph{ACM Computing Surveys,} 13(2):185-221, 1981.

\bibitem{!burrows:chubby}
Michael Burrows.
\newblock The Chubby Lock Service For {Loosely-Coupled} Distributed Systems.
\newblock In Proceedings of the \emph{Symposium on Operating System Design and Implementation,} pages 335-350, Seattle, Washington, November 2006.

\bibitem{DBLP:conf/vldb/ChamberlinGY81}
Donald D. Chamberlin, A. M. Gilbert, and Robert A. Yost.
\newblock A History Of System R And SQL/Data System.
\newblock In Proceedings of the \emph{International Conference on Very Large Data Bases,} pages 456-464, 1981.

\bibitem{DBLP:conf/osdi/ChangDGHWBCFG06}
Fay Chang, Jeffrey Dean, Sanjay Ghemawat, Wilson C. Hsieh, Deborah A. Wallach, Michael Burrows, Tushar Chandra, Andrew Fikes, and Robert Gruber.
\newblock Bigtable: A Distributed Storage System For Structured Data.
\newblock In Proceedings of the \emph{Symposium on Operating System Design and Implementation,} pages 205-218, Seattle, Washington, November 2006.

\bibitem{copysets}
Asaf Cidon, Stephen Rumble, Ryan Stutsman, Sachin Katti, John Ousterhout, and and Mendel Rosenblum.
\newblock Copysets: Reducing The Frequency Of Data Loss In Cloud Storage.
\newblock In Proceedings of the \emph{USENIX Annual Technical Conference,} pages 37--48, San Jose, California, June 2013.

\bibitem{!title=pnuts:journal=pvldb}
Brian F. Cooper, Raghu Ramakrishnan, Utkarsh Srivastava, Adam Silberstein, Philip Bohannon, Hans-Arno Jacobsen, Nick Puz, Daniel Weaver, and Ramana Yerneni.
\newblock PNUTS: Yahoo!'s Hosted Data Serving Platform.
\newblock In \emph{Proceedings of the VLDB Endowment,} 1(2):1277-1288, 2008.

\bibitem{!author=hsieh:title=spanner:booktitle=osdi}
James C. Corbett, Jeffrey Dean, Michael Epstein, Andrew Fikes, Christopher Frost, J. J. Furman, Sanjay Ghemawat, Andrey Gubarev, Christopher Heiser, Peter Hochschild, Wilson C. Hsieh, Sebastian Kanthak, Eugene Kogan, Hongyi Li, Alexander Lloyd, Sergey Melnik, David Mwaura, David Nagle, Sean Quinlan, Rajesh Rao, Lindsay Rolig, Yasushi Saito, Michal Szymaniak, Christopher Taylor, Ruth Wang, and Dale Woodford.
\newblock Spanner: Google's {Globally-Distributed} Database.
\newblock In Proceedings of the \emph{Symposium on Operating System Design and Implementation,} pages 261-264, Hollywood, California, October 2012.

\bibitem{!liskov:granola}
James Cowling and Barbara Liskov.
\newblock Granola: Low-Overhead Distributed Transaction Coordination.
\newblock In Proceedings of the \emph{USENIX Annual Technical Conference,} 2012.

\bibitem{!author=das:title=elastras:year=2013}
Sudipto Das, Divyakant Agrawal, and Amr El Abbadi.
\newblock ElasTraS: An Elastic, Scalable, And Self-Managing Transactional Database For The Cloud.
\newblock In \emph{ACM Transactions on Database Systems,} 38(1):5, 2013.

\bibitem{!vogels:dynamo}
Giuseppe DeCandia, Deniz Hastorun, Madan Jampani, Gunavardhan Kakulapati, Avinash Lakshman, Alex Pilchin, Swaminathan Sivasubramanian, Peter Vosshall, and Werner Vogels.
\newblock Dynamo: Amazon's Highly Available Key-Value Store.
\newblock In Proceedings of the \emph{Symposium on Operating Systems Principles,} pages 205-220, Stevenson, Washington, October 2007.

\bibitem{!escriva:hyperdex}
Robert Escriva, Bernard Wong, and Emin G{\"u}n Sirer.
\newblock HyperDex: A Distributed, Searchable Key-Value Store.
\newblock In Proceedings of the \emph{SIGCOMM Conference,} pages 25-36, Helsinki, Finland, August 2012.

\bibitem{!flavio:transactional}
Daniel G{\'o}mez Ferro, Flavio Junqueira, Benjamin Reed, and Maysam Yabandeh.
\newblock Lock-Free Transactional Support For Distributed Data Stores.
\newblock Poster Session.  Symposium on Operating Systems Principles, Cascais, Portugal, 2011.

\bibitem{linear2pc}
M. J. Flynn, J. N. Gray, A. K. Jones, K. Lagally, H.  Opderbeck, G. J. Popek, B. Randell, J. H. Saltzer, and H. R. Wehle.
\newblock Notes On Data Base Operating Systems.
\newblock In Proceedings of the \emph{Operating Systems,} pages 469--471, 1978.

\bibitem{!author=geambasu:title=comet:booktitle=osdi}
Roxana Geambasu, Amit A. Levy, Tadayoshi Kohno, Arvind Krishnamurthy, and Henry M. Levy.
\newblock Comet: An Active Distributed Key-Value Store.
\newblock In Proceedings of the \emph{Symposium on Operating System Design and Implementation,} pages 323-336, Vancouver, Canada, October 2010.

\bibitem{!arvind:scatter}
Lisa Glendenning, Ivan Beschastnikh, Arvind Krishnamurthy, and Thomas E. Anderson.
\newblock Scalable Consistency In Scatter.
\newblock In Proceedings of the \emph{Symposium on Operating Systems Principles,} pages 15-28, Cascais, Portugal, October 2011.

\bibitem{!lamport:transaction}
Jim Gray and Leslie Lamport.
\newblock Consensus On Transaction Commit.
\newblock In \emph{ACM Transactions on Database Systems,} 31(1):133-160, 2006.

\bibitem{gribble}
Steven D. Gribble.
\newblock A Design Framework And A Scalable Storage Platform To Simplify Internet Service Construction.
\newblock PhD thesis, U.C. Berkeley, 2000.

\bibitem{zookeeper}
Patrick Hunt, Mahadev Konar, Flavio P. Junqueira, and Benjamin Reed.
\newblock ZooKeeper: Wait-Free Coordination For Internet-Scale Systems.
\newblock In Proceedings of the \emph{USENIX Annual Technical Conference,} 2010.

\bibitem{!madden:mdcc}
Tim Kraska, Gene Pang, Michael J. Franklin, and Samuel Madden.
\newblock MDCC: Multi-Data Center Consistency.
\newblock In \emph{The Computing Research Repository,} abs/1203.6049, 2012.

\bibitem{DBLP:journals/tods/KungR81}
H. T. Kung and John T. Robinson.
\newblock On Optimistic Methods For Concurrency Control.
\newblock In \emph{ACM Transactions on Database Systems,} 6(2):213-226, 1981.

\bibitem{cassandra}
Avinash Lakshman and Prashant Malik.
\newblock Cassandra:  A Decentralized Structured Storage System.
\newblock In Proceedings of the \emph{International Workshop on Large Scale Distributed Systems and Middleware,} Big Sky, Montana, October 2009.

\bibitem{!lamport:parliament}
Leslie Lamport.
\newblock The {Part-Time} Parliament.
\newblock In \emph{ACM Transactions on Computer Systems,} 16(2):133-169, 1998.

\bibitem{!author=lamport:title=ordering:year=1978}
Leslie Lamport.
\newblock Time, Clocks, And The Ordering Of Events In A Distributed System.
\newblock In \emph{Communications of the ACM,} 21(7):558-565, 1978.

\bibitem{!shapiro:crdts}
Mihai Letia, Nuno M. Pregui\c{c}a, and Marc Shapiro.
\newblock CRDTs: Consistency Without Concurrency Control.
\newblock In \emph{The Computing Research Repository,} abs/0907.0929, 2009.

\bibitem{!lomet:deuteronomy}
Justin J. Levandoski, David B. Lomet, Sudipta Sengupta, Ryan Stutsman, and Rui Wang.
\newblock High Performance Transactions In Deuteronomy.
\newblock In Proceedings of the \emph{Conference on Innovative Data Systems Research,} Asilomar, California, January 2015.

\bibitem{!lloyd:cops}
Wyatt Lloyd, Michael J. Freedman, Michael Kaminsky, and David G. Andersen.
\newblock Don't Settle For Eventual: Scalable Causal Consistency For Wide-Area Storage With {COPS.}
\newblock In Proceedings of the \emph{Symposium on Operating Systems Principles,} pages 401-416, Cascais, Portugal, October 2011.

\bibitem{!lloyd:stronger}
Wyatt Lloyd, Michael J. Freedman, Michael Kaminsky, and David G. Andersen.
\newblock Stronger Semantics For Low-Latency Geo-Replicated Storage.
\newblock In Proceedings of the \emph{Symposium on Networked System Design and Implementation,} Lombard, Illinois, April 2013.

\bibitem{!lomet:unbundling}
David B. Lomet, Alan Fekete, Gerhard Weikum, and Michael J. Zwilling.
\newblock Unbundling Transaction Services In The Cloud.
\newblock In Proceedings of the \emph{Conference on Innovative Data Systems Research,} Asilomar, California, January 2009.

\bibitem{rococo}
Shuai Mu, Yang Cui, Yang Zhang, Wyatt Lloyd, and Jinyang Li.
\newblock Extracting More Concurrency From Distributed Transactions.
\newblock In Proceedings of the \emph{Symposium on Operating Systems Principles,} Colorado, October 2014.

\bibitem{!agrawal:g-store}
Faisal Nawab, Vaibhav Arora, Divyakant Agrawal, and Amr El Abbadi.
\newblock Minimizing Commit Latency Of Transactions In {Geo-Replicated} Data Stores.
\newblock In Proceedings of the \emph{SIGMOD International Conference on Management of Data,} pages 1279-1294, Melbourne, Victoria, Australia, May 2015.

\bibitem{!author=ousterhout:title=ramcloud:booktitle=sosp}
Diego Ongaro, Stephen M. Rumble, Ryan Stutsman, John K. Ousterhout, and Mendel Rosenblum.
\newblock Fast Crash Recovery In {RAMCloud.}
\newblock In Proceedings of the \emph{Symposium on Operating Systems Principles,} pages 29-41, Cascais, Portugal, October 2011.

\bibitem{!dabek:distributed-transactions}
Daniel Peng and Frank Dabek.
\newblock Large-Scale Incremental Processing Using Distributed Transactions And Notifications.
\newblock In Proceedings of the \emph{Symposium on Operating System Design and Implementation,} pages 251-264, Vancouver, Canada, October 2010.

\bibitem{voldemort}
Project Voldemort.
\newblock \url{http://project-voldemort.com/.}

\bibitem{!ratnasamy:content-addressable}
Sylvia Ratnasamy, Paul Francis, Mark Handley, Richard M. Karp, and Scott Shenker.
\newblock A Scalable Content-Addressable Network.
\newblock In Proceedings of the \emph{SIGCOMM Conference,} pages 161-172, San Diego, California, August 2001.

\bibitem{riak}
Riak.
\newblock \url{http://basho.com/.}

\bibitem{pastry}
Antony I. T. Rowstron and Peter Druschel.
\newblock Pastry: Scalable, Decentralized Object Location, And Routing
               For Large-Scale Peer-To-Peer Systems.
\newblock In Proceedings of the \emph{IFIP/ACM International Conference on Distributed Systems Platforms,} pages 329-350, 2001.

\bibitem{scalaris}
Thorsten Sch\"{u}tt, Florian Schintke, and Alexander Reinefeld.
\newblock Scalaris: Reliable Transactional P2p Key/value Store.
\newblock In Proceedings of the \emph{SIGPLAN Workshop on ERLANG,} pages 41-48, Victoria, Canada, 2008.

\bibitem{sdur-geo}
Daniele Sciascia and Fernando Pedone.
\newblock Geo-Replicated Storage With Scalable Deferred Update Replication.
\newblock In Proceedings of the \emph{International Conference on Dependable Systems and Networks,} pages 1-12, Budapest, Hungary, June 2013.

\bibitem{sdur}
Daniele Sciascia, Fernando Pedone, and Flavio Junqueira.
\newblock Scalable Deferred Update Replication.
\newblock In Proceedings of the \emph{International Conference on Dependable Systems and Networks,} pages 1-12, Boston, Massachusetts, June 2012.

\bibitem{!sovran:transactional}
Yair Sovran, Russell Power, Marcos K. Aguilera, and Jinyang Li.
\newblock Transactional Storage For Geo-Replicated Systems.
\newblock In Proceedings of the \emph{Symposium on Operating Systems Principles,} pages 385-400, Cascais, Portugal, October 2011.

\bibitem{!stoica:chord}
Ion Stoica, Robert Morris, David R. Karger, M. Frans Kaashoek, and Hari Balakrishnan.
\newblock Chord: A Scalable Peer-To-Peer Lookup Service For Internet Applications.
\newblock In Proceedings of the \emph{SIGCOMM Conference,} pages 149-160, San Diego, California, August 2001.

\bibitem{!stonebraker:complete-rewrite}
Michael Stonebraker, Samuel Madden, Daniel J. Abadi, Stavros Harizopoulos, Nabil Hachem, and Pat Helland.
\newblock The End Of An Architectural Era (It's Time For A Complete Rewrite).
\newblock In Proceedings of the \emph{International Conference on Very Large Data Bases,} pages 1150-1160, 2007.

\bibitem{bocc}
H\"{a}rder Theo.
\newblock Observations On Optimistic Concurrency Control Schemes.
\newblock In \emph{Information Systems,} 9(2):111-120, 1984.

\bibitem{!abadi:calvin}
Alexander Thomson, Thaddeus Diamond, Shu-Chun Weng, Kun Ren, Philip Shao, and Daniel J. Abadi.
\newblock Calvin: Fast Distributed Transactions For Partitioned Database Systems.
\newblock In Proceedings of the \emph{SIGMOD International Conference on Management of Data,} pages 1-12, Scottsdale, Arizona, May 2012.

\bibitem{!schneider:chain}
Robbert van Renesse and Fred B. Schneider.
\newblock Chain Replication For Supporting High Throughput And Availability.
\newblock In Proceedings of the \emph{Symposium on Operating System Design and Implementation,} pages 91-104, San Francisco, California, December 2004.

\bibitem{lynx}
Yang Zhang, Russell Power, Siyuan Zhou, Yair Sovran, Marcos K.  Aguilera, and Jinyang Li.
\newblock Transaction Chains: Achieving Serializability With Low Latency In Geo-Distributed Storage Systems.
\newblock In Proceedings of the \emph{Symposium on Operating Systems Principles,} Pennsylvania, November 2013.

\bibitem{!kubiatowicz:tapestry}
Ben Y. Zhao, John Kubiatowicz, and Anthony D. Joseph.
\newblock Tapestry: A Fault-Tolerant Wide-Area Application Infrastructure.
\newblock In \emph{SIGCOMM Computer Communications Review,} 32(1):81, 2002.

\end{thebibliography}

\end{document}